\documentclass[%
 reprint,
 amsmath,amssymb,
 aps,
 prl
]{revtex4-2}

\usepackage{graphicx}
\usepackage{dcolumn}
\usepackage{bm}
\usepackage{xcolor}
\usepackage[bottom]{footmisc}

 \newcommand{\bfB}{{\bf{B}}}
  \newcommand{\bfr}{{\bf{r}}}
   \newcommand{\bfm}{{\bf{m}}}
   \newcommand{\bfn}{{\bf{n}}}

\usepackage{caption}
\usepackage{subcaption}
\begin{document}


\title{A New Degree of Freedom in Stellarator Coil Design}

\author{Sophia A. Henneberg}
\affiliation{Plasma Science and Fusion Center, Massachusetts Institute of Technology, Cambridge, MA 02139, USA}%
\affiliation{Max-Planck-Institut f\"{u}r Plasmaphysik, Wendelsteinstr. 1, 17489 Greifswald}%


\author{Annika Zettl}
\affiliation{Max-Planck-Institut f\"{u}r Plasmaphysik, Wendelsteinstr. 1, 17489 Greifswald}%


\date{\today}

\begin{abstract}
Reversed toroidal field (RTF) coils introduce a new degree of freedom in stellarator coil design. While the total plasma‑linked current is fixed by Ampère’s law, compensating RTF currents allow the modular‑coil current to exceed this value without altering the net linkage. This redistribution decouples key coil constraints and modifies the external coil geometry without significantly compromising the confining magnetic field. As a result, the achievable coil–plasma separation can be increased in the configurations studied, potentially enabling more compact stellarator designs.
\end{abstract}

\maketitle

Engineering constraints play a central role in determining the feasibility and minimum size of magnetic confinement fusion devices \cite{Najmabadi-2008, Lion-2025, Hegna-2025}. A key limitation is the required coil–plasma separation needed to accommodate the breeding blanket and neutron shielding. Although this constraint can be met by uniformly scaling devices to larger size, such scaling suffers substantial cost penalties. Achieving reactor‑relevant performance at smaller size therefore offers a compelling pathway to reduced capital cost \cite{Najmabadi-2008}.
Numerous studies have sought to increase the coil–plasma separation, either by incorporating coil geometry directly into the optimization \cite{Henneberg-2021, Giuliani-2022, Giuliani-2024} or by developing improved proxies for this distance \cite{Drevlak-2019, Henneberg-2019, Kappel-2024, Kappel-2026}. While the latter approach has recently demonstrated substantial gains \cite{Goodman-2024, Schuett-2025}, these improvements typically come at the expense of other favorable design properties.

Here, we present an approach that can further increase the coil–plasma separation and can be applied independently or in combination with existing methods. Conventional stellarator coil configurations, such as W7‑X \cite{Geiger-2015}, employ nonplanar, plasma‑linked modular coils. By contrast, there are also concepts based on permanent magnets \cite{Helander-2020} or dipole coils with simplified toroidal field coils \cite{Gates-2025, Kaptanoglu-2025}. The latter employ unlinked, typically planar coils, offering greater flexibility in current distribution and spacing. We demonstrate that augmenting a standard modular coil set with additional plasma‑linked reversed toroidal field coils introduces a comparable degree of current‑space freedom, potentially enabling simplified coil geometries and increased coil–plasma separation.



\begin{figure}[htb]
  \centering
  \includegraphics[width=0.18\textwidth]{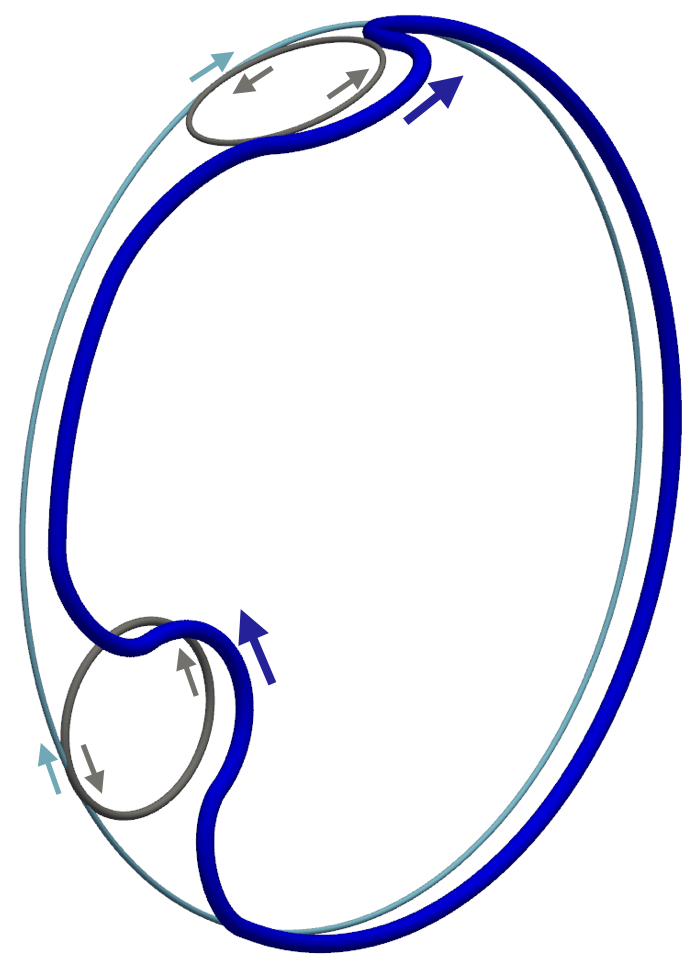}
  \caption{A modular coil (dark blue) can be decomposed into a large planar toroidal field (TF) coil (light blue) and a set of smaller dipole coils (gray). The dipoles may be oriented parallel to the plasma boundary (top), perpendicular to it (bottom), or in a combination of both.}
  \label{fig:dipoles}
\end{figure}

Modular coils typically have complex, three‑dimensionally curved shapes, with their magnetic field dominated by the toroidal component. They can be approximately represented by a superposition of a planar, plasma‑linked toroidal field (TF) coil, and a set of smaller dipole coils, as illustrated in Fig \ref{fig:dipoles}. Increasing the number of dipole coils while reducing their individual size improves this approximation. Adjacent coil elements carry oppositely directed currents, leading to local cancellation such that the resulting effective current distribution closely traces the modular coil geometry. Note, this decomposition into toroidal and dipole components is a conceptual tool for deriving our magnetic field scaling; it does not imply a physical separation of the modular coils.


 By conceptually decomposing modular coils into TF and dipole components, each contribution can be analyzed independently. In particular, the magnetic field generated by a dipole coil, at distances large compared to its radius, approaches that of an ideal magnetic dipole,
     \begin{equation}
         \bfB_\text{dipole}(\bfr)= \frac{\mu_0}{4\pi r^5} (3\bfr(\bfm\cdot\bfr)-\bfm r^2),
     \end{equation}
      where $\mu_0$ is the vacuum permeability, $\bfr$ is the distance vector from the dipole, $\bfm=I\,A\,\bfn$ is the magnetic moment defined by the coil current, $I$, enclosed area $A$, and the normal vector of the area $\bfn$. Asymptotically, the dipole magnetic field scales as $B_\text{dipole}\sim I/r^3$ from afar, implying that the field strength can be maintained at larger coil–plasma distances by increasing the current according to $I\sim r^3$. 
    If the coil current can be adjusted, the dipole components can in principle be placed farther from the plasma by increasing their current. However, when dipole and TF components are part of the same modular coil, the dipole current is constrained by the TF current, which is itself effectively fixed. For a given equilibrium, the total current linked with the plasma, $I_{tot}$
    , is determined by the plasma properties via Ampère’s law,
    \begin{equation}
        \mu_0 I_\text{tot} = \oint_\mathcal{C} \mathbf{B}\cdot \text{d}\mathbf{l},
    \end{equation}
    where $\mathcal{C}$ is a closed, non-linked contour on the plasma boundary with the line element $\text{d}\mathbf{l}$ along $\mathcal{C}$, and $\mathbf{B}$ is the magnetic field.
     In standard stellarator optimizations, the modular coils provide the sole source of linked current, and their total current is therefore constrained to satisfy $I_\text{tot}=I_\text{MC}$.
    By adding a ``zero” current, $I_{\text{zero}}=0$, to the total plasma‑linked current, the total modular‑coil current can be varied without altering the equilibrium, formally written as $I_{\text{tot}} = I_{\text{tot}} + I_{\text{zero}}$. Choosing $I_{\text{zero}} = I_{\text{add}} + I_{\text{RTF}}$, where $I_{\text{RTF}}$ denotes the total current in auxiliary RTF coils, allows the modular‑coil current to be increased by an additional amount $I_{\text{add}}$, such that $I_{\text{MC}} = I_{\text{tot}} + I_{\text{add}}$. In this way, the effective current constraint on the modular coils can be overcome.

A complete description of the magnetic field generated by modular coils requires more than the dipole approximation alone. The contribution from the TF coil component can be estimated to leading orders by the magnetic field at its center, 
$B_{\text{TF}} = \mu_0 I \ (1/2R + 3x^2/8R^3)$,
where $R$ is the TF coil radius, and $x$ denotes a small distance from the center ($x\ll R$),
assuming a coil–plasma separation $r \sim R$. For sufficiently large separations $r$, the total magnetic field produced by the modular coils is therefore expected to scale as
\begin{equation}
B_{\text{MC}}(r) \sim B_{\text{dipole}} + B_{\text{TF}}
\sim c_{\text{mixed}}\,\frac{I}{r^{3}} + c_{\text{TF}}\,\frac{I}{r},
\label{eq:modcoilscaling}
\end{equation}
where $c_{\text{mixed}}$ and $c_{\text{TF}}$ are geometry‑dependent coefficients. It then follows that the required current scales with distance as $I \sim \left(c_{\text{mixed}}/r^{3} + c_{\text{TF}}/r\right)^{-1}$.

The coils are optimized using a ``stage‑II" approach, in which the plasma boundary is fixed and the normal magnetic field error (along with additional objectives) is minimized to obtain coils that approximately reproduce the prescribed boundary \cite{Henneberg-2021}. All coils are modeled as filamentary currents, an appropriate approximation when the coil–plasma separation is large compared to the coil thickness. The optimizations are carried out using the coil optimization framework in \texttt{simsopt} \cite{Landreman-2021}.
The optimized coil configuration consists of two coil types. There are $N$ distinct modular coils per half field period, which repeat around the torus consistent with stellarator symmetry. Unless otherwise noted, modular‑coil optimizations employ a cold‑start initialization, in which the coil shapes are initially planar and circular and have not been previously optimized. In addition, 20 RTF coils are placed uniformly around the torus, surrounding the modular coils. For this initial study, the RTF coils are circular planar coils and do not change in the optimization. Since RTF coil shape and position affect the modular coil optimization outcome, we maintain these parameters constant to ensure clean scaling. Optimization of the RTF coil geometry is deferred to future work.
  The total plasma‑linked current, $I_{\text{tot}}$, is fixed by the target equilibrium, as described above, and is divided between the modular coils and the RTF coils according to $I_{\text{tot}} = I_{\text{MC}} + I_{\text{RTF}}$. We define the current ratio as the fraction of the total linked current carried by the modular coils,
$Q_{\text{c}} \equiv I_{\text{MC}}/I_{\text{tot}} = 1 + I_{\text{add}}/I_{\text{tot}}$.
For simplicity, the currents $I_{\text{MC}}$ and $I_{\text{RTF}}$ are distributed uniformly among the modular and RTF coils, respectively, and all individual coil currents are held fixed during optimization. Consequently, only the geometric shapes of the modular coils are varied.
    For a range of current ratios, $0.5 \leq Q_{\text{c}} \leq 2$, we optimize multiple coil sets for boundaries corresponding to two distinct configurations: a compact stellarator (major radius $2.5\,\mathrm{m}$, aspect ratio $5$, $n_{\mathrm{fp}}=4$, $I_\text{tot} = 45.6\,\mathrm{MA}$, quasi‑axisymmetric, $\beta=1.2\%$) \cite{Henneberg-2024, Schuett-2024}, and the W7‑X standard configuration (major radius $5.5\,\mathrm{m}$, aspect ratio $10.75$, $n_{\mathrm{fp}}=5$, $I_\text{tot} = 81.0\,\mathrm{MA}$, quasi‑isodynamic, vacuum field) \cite{Geiger-2015}.
    The optimization targets are deliberately chosen to be minimal and initially exclude any explicit dependence on the coil–plasma distance. For the compact stellarator, the objective function includes three terms: the quadratic flux error, $F_{\Phi}$, the total coil length, $F_{\mathrm{L}}$, and a minimum coil–coil separation penalty, $F_{\mathrm{cc}}$, yielding
$J_\text{QA} = F_{\Phi} + F_\text{L} + F_\text{cc}$.
For the W7‑X configuration, additional constraints on the local curvature, $F_{\kappa}$, and the mean‑squared curvature, $F_{\mathrm{msc}}$, are required to obtain feasible coil sets, resulting in
$J_\text{W7-X} = F_{\Phi} + F_\text{L} + F_\text{cc} + F_{\kappa} + F_\text{msc}$.
To examine achievable coil–plasma separations, we perform additional optimizations in which the target function explicitly includes a coil–surface distance objective, $F_{\text{cs}}$, designed to maximize the separation from the plasma boundary. All objectives are formulated using standard metrics commonly employed in coil optimization studies (see, e.g., Ref.~\cite{hurwitz2025electromagnetic}). Both weights and target values were held fixed throughout each current ratio scan.

\begin{figure}[htb]
  \centering
  \begin{subfigure}[b]{0.5\textwidth}
  \centering
    \includegraphics[width=1\textwidth]{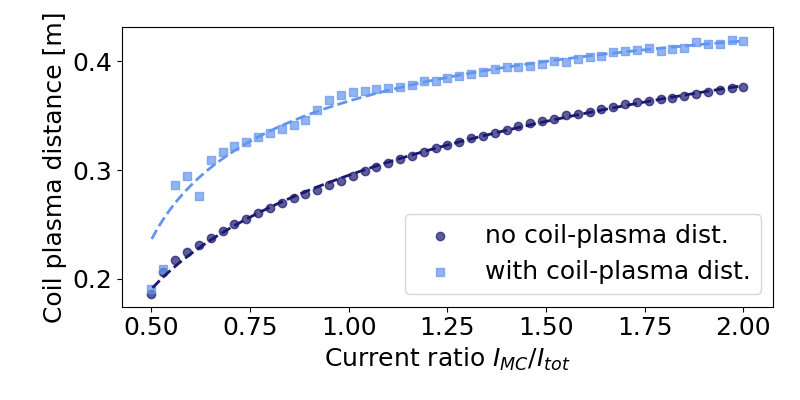}
    \caption{Dark blue: coils optimized for field error, length, and coil–coil separation; light blue: additionally optimized for a $0.6\,\mathrm{m}$ coil–plasma distance.}
    \label{fig:hybrid_CStarget}
  \end{subfigure}
  \begin{subfigure}[b]{0.5\textwidth}
  \centering
      \includegraphics[width=1\textwidth]{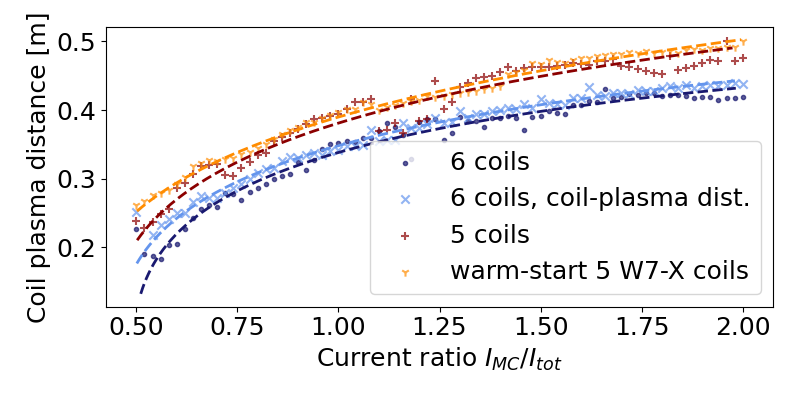}
      \caption{Blue: six modular coils; red/orange: five. Light blue additionally targets a $0.6\,\mathrm{m}$ coil–plasma distance. The largest separations arise from warm‑start optimizations of the original W7‑X coils (orange), reproducing the $Q_{\text{c}}=1$ target.}
      \label{fig:W7-warmstart}
  \end{subfigure}
  \caption{Coil–plasma separation as a function of the current ratio, $Q_{\text{c}} = I_{\text{MC}}/I_{\text{tot}}$, for coils optimized for (a) a compact stellarator and (b) the W7‑X standard configuration. 
  Dashed lines show fits to Eq.~\ref{eq:currentdistance}.}
\end{figure}

    Motivated by Eq.~\ref{eq:modcoilscaling}, we expect the current ratio to relate to the coil-plasma distance as
   \begin{equation}
       Q_\text{c} =\left( \frac{c_\text{mixed}}{r^{3}}+\frac{c_\text{TF}}{r}\right)^{-1} + c_{I},
       \label{eq:currentdistance}
   \end{equation}
   where $c_{I}$ is an additional constant, which is necessary to encompass that at a hypothetical zero coil-plasma distance, there is a non-zero current necessary in the modular coils to generate the dipole field components.
   The coefficients $c_{\text{mixed}}$, $c_{\text{TF}}$, and $c_I$ are fitted to the optimization scans for both equilibria, yielding good agreement with the numerical results (Figs.~\ref{fig:hybrid_CStarget} and \ref{fig:W7-warmstart}).
   
   For the compact stellarator, doubling the modular‑coil current results in a $27.8\%$ increase in the minimum coil–plasma separation, see Fig.~\ref{fig:hybrid_CStarget}. Notably, this trend persists even when the coils are explicitly optimized for coil–plasma distance, in which case the minimum distance remains $12.6\%$ larger at $Q_{\text{c}}=2$. In this optimization, we also observe simultaneous reductions in total coil length and curvature with the addition of RTF coils. Similarly, for the cold-start W7‑X optimization, we find an increase in minimum coil–plasma distance of $19$–$28\%$, both with and without an explicit distance objective, see Fig.~\ref{fig:W7-warmstart}.
   
\begin{figure}[htb]
  \centering
  \includegraphics[width=0.5\textwidth]{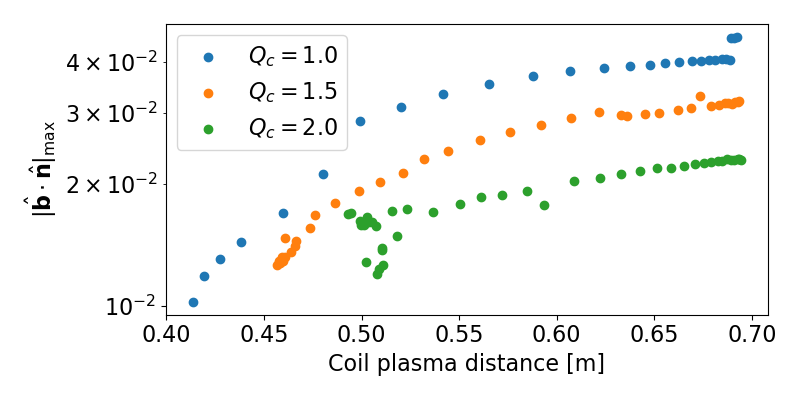}
  \caption{Maximum absolute normal magnetic field, $|\hat{\mathbf{b}}\cdot\hat{\mathbf{n}}|_{\max}$, where $\hat{\mathbf{b}}=\mathbf{B}/B$, versus coil–plasma distance with five modular coils per half field period and RTF coils for W7‑X. Current ratio: $Q_{\text{c}}=1$ (blue), $1.5$ (orange), and $2$ (green). 
  }
  \label{fig:W7-CSweight_scan}
\end{figure}

   Typically, the field error $|\vec{B}\cdot \hat{n}|$ increases  when enforcing a higher coil-plasma distance in the optimization. However, adding RTF coils can increase the distance while maintaining the same field error. That means that with a higher current ratio, it is easier to achieve lower normal field errors for the same coil-plasma distance. This relationship is illustrated in Figure \ref{fig:W7-CSweight_scan}, where larger separations are achieved by increasing the distance weight, targeting a minimum distance of $0.7\,\mathrm{m}$. Note that with each current ratio, there also appears to be some 'natural' coil-plasma distance, where the field error is lowest in the absence of an explicit distance objective.
   
Because the attainable coil–plasma separation increases with coil current, the number of modular coils plays an important role. For W7‑X, adding a sixth modular coil to the standard five‑coil configuration requires a reduction in the current in each coil by one sixth ($\sim17\%$), which is reflected in the optimizations by systematically smaller coil–plasma distances. Notably, for five coils the normal field error increases slightly with increasing current ratio, whereas for six coils the optimizations readily identify configurations with even smaller flux errors. This behavior suggests the existence of an optimal number of coils for a given coil–plasma separation. Consequently, when increasing the modular‑coil current to achieve larger separation, it may—in certain cases, as here—be advantageous to introduce an additional coil.
   
\begin{figure}[h]
  \centering
  \begin{subfigure}[b]{0.4\textwidth}
  \centering
    \includegraphics[width=1\textwidth]{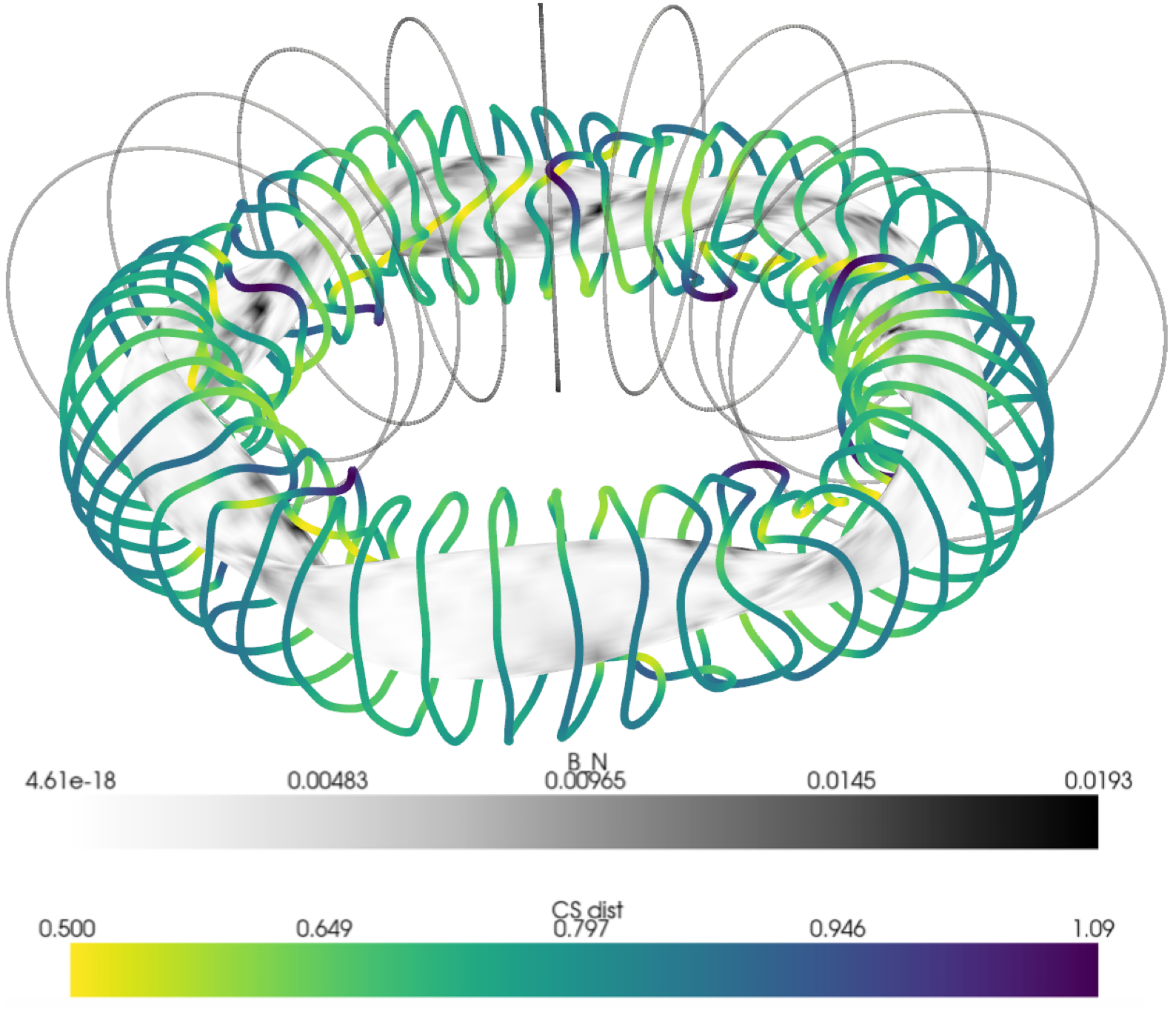}
    \caption{Modular coils with 10 out of 20 total Reversed Toroidal Field coils for visual clarity. The gray-scale indicates the normal field error, and the color scale the coil-plasma distance.}
    \label{fig:W7-CQ2coils}
  \end{subfigure}
  \begin{subfigure}[b]{0.4\textwidth}
  \centering
      \includegraphics[width=0.6\textwidth]{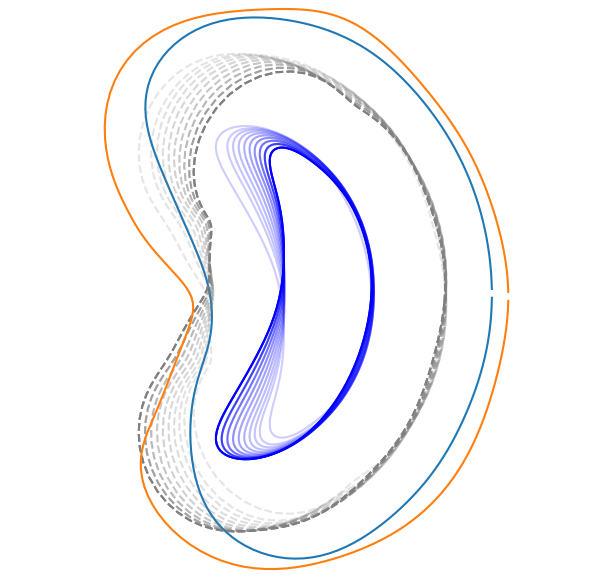}
      \caption{Toroidal projections of coils near the bean cross‑section for the original W7‑X coils (light blue) and the new modular coils (orange). Contours show ten consecutive toroidal cross‑sections of the plasma boundary (dark blue) and a $0.4\,\mathrm{m}$ plasma offset (gray), spaced by $2^\circ$.
      }
  \end{subfigure}
  \caption{Coils and plasma boundary of W7-X with current ratio 2.}
  \label{fig:crosssection}
\end{figure}

Finally, we repeat the optimization using a `warm-start' based on the existing W7‑X coil set, which was originally optimized without RTF coils. This approach yields larger coil–plasma separation while preserving comparable coil‑quality metrics. When the W7‑X coils are re‑optimized together with RTF coils, the coil–plasma distance increases by more than $25\%$ at a current ratio of $Q_{\text{c}}=2$ (orange curve in Fig.~\ref{fig:W7-warmstart}). The resulting coil geometry, illustrating the increased separation, is shown in Fig.~\ref{fig:crosssection}.


 The inclusion of Reversed Toroidal Field coils introduces additional freedom in the choice of modular‑coil current while preserving the total plasma‑linked current. We demonstrate that this freedom could be exploited to increase the coil–plasma separation.
    
In the past, some designs involved planar toroidal field coils in addition to the modular coils, e.g. Wendelstein 7-X \cite{Wanner-2001} or NCSX \cite{Nelson-2003}. However, these toroidal field coils typically have been added after the design of the modular coils for increasing experimental flexibility \cite{Wanner-2001}. 
In other cases, it was included in the design process but divided the total linking current between toroidal field coils and modular coils \textit{without} reversing the direction.
This choice could worsen the engineering properties (except the coil current) of the modular coils -- from our analysis, this is not surprising at all.
    
Our results are consistent with the findings of Kappel et al. based on the $L_{\nabla B}$ metric \cite{Kappel-2024}. That work demonstrated that the maximum current density on a current‑potential surface significantly influences the attainable coil–plasma distance (see Fig.~4 of Ref.~\cite{Kappel-2024}). The reported trend is also well captured by the scaling relation in Eq.~\ref{eq:currentdistance}.

One might expect that increasing the toroidal magnetic field would ease modular‑coil constraints by raising the total plasma‑linked current via Ampère’s law. However, this is not the case when the rotational transform, $\iota$, and aspect ratio are held fixed. Under these conditions, the toroidal and poloidal magnetic fields scale proportionally, so an increase in the toroidal field necessarily produces a corresponding linear increase in the poloidal field, leaving the overall constraint unchanged.

Note that the reported coil-plasma distances do not include the finite-build extent of the coils. One might expect that the coil thickness would increase with larger currents, scaling as $\sim\sqrt{Q_c}$. Actual values depend on the choice of conductor material and coil design which is left for future work. 

As an alternative to increasing the coil–plasma separation, the dipole strength in the decomposed modular‑coil model can be modified through $B_{\text{dipole}} \sim A I$. Reducing the current therefore requires a compensating increase in the dipole area $A$ to preserve the magnetic field. In practice, this corresponds to increased coil non‑planarity, leading to more complex geometries and degraded curvature properties.


We emphasize that the present work aims to introduce the concept of reversed toroidal field coils. This opens up the possibility of future studies which may extend this approach by optimizing the RTF coil shapes, positions, and currents to further improve engineering performance, including electromagnetic forces. Preliminary investigations indicate that forces on the coils can increase, slightly decrease, or remain unchanged, suggesting substantial scope for improvement through optimization. In addition, one could envision deploying only a limited number of RTF coils at locations where modular coils approach the plasma most closely, thereby reducing the number of additional coils and current while retaining the benefits demonstrated here.
      
In summary, the inclusion of reversed toroidal field coils adds an additional degree of freedom to stellarator optimization, enabling either increased coil–plasma separation or simplified modular‑coil geometries. Both effects offer potential pathways to reduced device size and lower engineering complexity, with corresponding cost benefits.

\begin{acknowledgments}
This work has been carried out within the framework of the EUROfusion Consortium, funded by the European Union via the Euratom Research and Training Programme (Grant Agreement No 101052200 — EUROfusion). Views and opinions expressed are however those of the author(s) only and do not necessarily reflect those of the European Union or the European Commission. Neither the European Union nor the European Commission can be held responsible for them.\\
This research was supported by the MIT Center for International Studies’ Global Seed Fund program.\\
This work was supported by the U.S. Department of Energy contract number DE-FG02-91ER-54109.

\end{acknowledgments}

\appendix



\bibliography{sophia}

\end{document}